\documentclass[twocolumn,prl,showpacs,amsfonts,amsmath,assymb,eufrak,superscriptaddress]{revtex4-1}
\usepackage{amsmath,amsfonts,amssymb,mathtools,upgreek}
\usepackage{epsfig}

\usepackage[dvipdfmx]{hyperref}\hypersetup{hidelinks}

\def\eq#1\en{\begin{equation}#1\end{equation}} 
\def\eqa#1\ena{\begin{align}#1\end{align}}
\def\eqg#1\eng{\begin{gather}#1\end{gather}}
\newcommand{\lb}[1]{\label{e:#1}}
\newcommand{\rlb}[1]{\eqref{e:#1}} 
\newcommand{\nl}{\notag\\}

\newcommand{\norms}[1]{\Vert#1\Vert}

\newcommand{\sumtwo}[2]%
{\mathop{\sum_{#1}}_{#2}}
\newcommand{\sumthree}[3]%
{\mathop{\mathop{\sum_{#1}}_{#2}}_{#3}}
\newcommand{\sumfour}[4]%
{\mathop{\mathop{\mathop{\sum_{#1}}_{#2}}_{#3}}_{#4}} 
\newcommand{\prodtwo}[2]%
{\mathop{\prod_{#1}}_{#2}}
\newcommand{\mintwo}[2]%
{\mathop{\min_{#1}}_{#2}}
\newcommand{\maxtwo}[2]%
{\mathop{\max_{#1}}_{#2}}
\newcommand{\maxthree}[3]%
{\mathop{\mathop{\max_{#1}}_{#2}}_{#3}}
\newcommand{\limtwo}[2]%
{\mathop{\lim_{#1}}_{#2}}
\newcommand{\suptwo}[2]%
{\mathop{\sup_{#1}}_{#2}}
\newcommand{\supthree}[3]%
{\mathop{\mathop{\sup_{#1}}_{#2}}_{#3}}
\newcommand{\supfour}[4]%
{\mathop{\mathop{\mathop{\sup_{#1}}_{#2}}_{#3}}_{#4}} 
\newcommand{\inftwo}[2]%
{\mathop{\inf_{#1}}_{#2}}
\newcommand{\infthree}[3]%
{\mathop{\mathop{\inf_{#1}}_{#2}}_{#3}}
\newcommand{\inffour}[4]%
{\mathop{\mathop{\mathop{\inf_{#1}}_{#2}}_{#3}}_{#4}} 

\newcommand{\bsr}{\boldsymbol{r}}
\newcommand{\bss}{\boldsymbol{s}}

\newcommand{\bbC}{\mathbb{C}}

\newcommand{\bbR}{\mathbb{R}}

\newcommand{\up}{\uparrow}
\newcommand{\dn}{\downarrow}

\newcommand{\Di}{\mathit{\Delta}}
\newcommand{\La}{\Lambda}
\newcommand{\qedm}{\rule{1.5mm}{3mm}}

\newcommand{\Lap}{\boldsymbol{\bigtriangleup}}
\newcommand{\bra}[1]{\langle#1|}
\newcommand{\ket}[1]{|#1\rangle}

\newcommand{\hS}{\hat{S}}
\newcommand{\hH}{\hat{H}}
\newcommand{\hN}{\hat{N}}
\newcommand{\hn}{\hat{n}}
\newcommand{\ha}{\hat{a}}
\newcommand{\had}{\hat{a}^\dagger}

\newcommand{\Gk}{\ket{\Phi^{\rm GS}_N}}
\newcommand{\Gb}{\bra{\Phi^{\rm GS}_N}}

\newcommand{\NLa}{|\La|}

\newcommand{\EGS}{E^{\rm GS}}

\newcommand{\hO}{\hat{\mathcal{O}}}
\newcommand{\hOd}{\hat{\mathcal{O}}^\dagger}
\newcommand{\bkP}[1]{\Gb#1\Gk}
\newcommand{\bkG}[1]{\langle#1\rangle_{\!N}^{\!\rm GS}}
\newcommand{\bbkG}[1]{\bigl\langle#1\bigr\rangle_{\!N}^{\!\rm GS}}
\newcommand{\bkGS}[1]{\langle#1\rangle^{\!\rm GS}}

\newcommand{\hpsi}{\hat{\psi}}
\newcommand{\hpsid}{\hat{\psi}^\dagger}
\newcommand{\hh}{\hat{h}}
\newcommand{\ho}{\hat{o}}
\newcommand{\eGS}{\epsilon}

\newcommand{\tGk}{\ket{\tilde{\Phi}^{\rm GS}_{\mu,\sbf}}}
\newcommand{\tGb}{\bra{\tilde{\Phi}^{\rm GS}_{\mu,\sbf}}}
\newcommand{\tEGS}{\tilde{E}^{\rm GS}}
\newcommand{\tNGS}{\tilde{N}^{\rm GS}}
\newcommand{\sbf}{\lambda}

\newcommand{\para}[1]{\medskip\par{\em #1}\/.---}
\newcommand{\prop}[1]{{\em #1}\/.---}

\begin{document}
\title{Off-Diagonal Long-Range Order Implies Vanishing Charge Gap}

\author{Hal Tasaki}\email[]{hal.tasaki@gakushuin.ac.jp}
\affiliation{Department of Physics, Gakushuin University, Mejiro, Toshima-ku, Tokyo 171-8588, Japan}

\author{Haruki Watanabe} \email[]{hwatanabe@g.ecc.u-tokyo.ac.jp}
\affiliation{Department of
Applied Physics, University of Tokyo, Tokyo 113-8656, Japan.}

\date{\today}

\begin{abstract}
For a large class of quantum many-body systems with U(1) symmetry, we prove a general inequality that relates the (off-diagonal) long-range order with the charge gap.
For a system of bosons or fermions on a lattice or in the continuum, the inequality implies that a ground state with off-diagonal long-range order inevitably has a vanishing charge gap, and hence is characterized by nonzero charge susceptibility.
For a quantum spin system, the inequality implies that a ground state within a magnetization plateau cannot have transverse long-range order.
(There is a 16 minutes video in which the main results of the paper are described.  See \url{https://youtu.be/fvZdgV8Ik-8})
\end{abstract}


\maketitle

Equal-time correlation functions and energy gaps of a quantum many-body system are intimately connected.
The connection, which is trivial in relativistic field theories, is widely observed in non-relativistic theories as well, and has partly been established rigorously by means of the variational principle \cite{HorschLinden,KomaTasaki1994,SannomiyaKatsuraNakayama2017,Tasaki2017} or the Lieb-Robinson bound \cite{HastingsKoma,NachtergaeleSims}.
See also \cite{Wagner1966,Stringari,Momoi94,Momoi95,Koma2021} for rigorous results for non-relativistic systems related to the Goldstone theorem.

In this Letter, we prove, in a large class of macroscopic quantum many-body systems with U(1) symmetry, a general inequality that relates the order parameter for (off-diagonal) long-range order in the ground state with the charge gap.
When applied to a system of bosons or fermions on a lattice or in the continuum, our inequality implies that a ground state with nonzero off-diagonal long-range order (ODLRO) is inevitably accompanied by a vanishing charge gap, or, almost equivalently, characterized by nonzero charge susceptibility $\chi(\rho)=\{d\mu(\rho)/d\rho\}^{-1}$, where $\mu(\rho)$ is the chemical potential and $\rho$ the particle density.
This means that the ground state with ODLRO is ``compressible'' when the chemical potential is varied.
This conclusion may be most meaningful (and even surprising) when applied to a commensurate supersolid \cite{OhgoeSuzukiKawashima2012,SuzukiKoga2014}.
See Discussion.
Conversely, the inequality implies that a ground state accompanied by a nonzero charge gap cannot exhibit ODLRO.
When applied to a quantum spin system, our inequality implies that a ground state within a magnetization plateau cannot have long-range order in the direction perpendicular to the magnetic field.
We note that these conclusions do not follow from the Goldstone theorem, which does not deal with the charge gap.

The theorem is proved by a simple variational argument that applies to a wide class of quantum many-body systems.
The proof was inspired to us by a series of closely related works on low-lying excited states in quantum spin systems whose exact ground states have long-range order but do not exhibit spontaneous symmetry breaking \cite{HorschLinden,KomaTasaki1994,SannomiyaKatsuraNakayama2017,Tasaki2017}.
See Part~I of \cite{TasakiBook} for a detailed account of the subject.

For lattice boson models with hardcore interaction, the main inequality \rlb{main} was proved by Tian \cite{Tian}.
(See also \cite{ALSSY} for the hardcore boson models at half-filling, which indeed reduces to spin systems \cite{HorschLinden,KomaTasaki1994,SannomiyaKatsuraNakayama2017,Tasaki2017,TasakiBook}.)
An inequality for the charge susceptibility that resembles our \rlb{kge} was proved in \cite{Wagner1966,Stringari} for a general class of models.
This inequality however is meaningful only in a ground state that breaks the U(1) symmetry, as we explain below.
As far as we know our result for quantum spin systems is new.

\para{Lattice boson systems}%
Let us state and prove our main result in the specific setting of lattice boson systems.
Generalizations are discussed below.

Let $\La$ be an arbitrary finite lattice, i.e., a set of sites $x,y,\ldots$, and $\NLa$ be the number of sites.
Consider a system of bosons on $\La$.
We denote by $\ha_x$, $\had_x$, and $\hn_x=\had_x\ha_x$ the annihilation, creation, and number operators at site $x$.
The total number operator is given by $\hN=\sum_{x\in\La}\hn_x$.
We sometimes refer to $\hN$ as the conserved charge.
Although we fix $\La$ in the following discussion, we always have in mind the statistical mechanical setting where we take $\NLa$ large and examine how various quantities scale with $\NLa$.

We consider the  general Hamiltonian
\eq
\hH=\sum_{x,y\in\La}t_{x,y}\,\had_x\ha_y+\sum_{x\in\La}\frac{U_x}{2}\hn_x(\hn_x-1)+\sum_{x,y\in\La}\frac{V_{x,y}}{2}\hn_x\hn_y,
\lb{H}
\en
where $t_{x,y}=t_{y,x}^*\in\bbC$ is the hopping amplitude, $U_x\in\bbR$ is the on-site repulsive energy, and $V_{x,y}=V_{y,x}\in\bbR$ with $V_{x,x}=0$ is the interaction energy.
Note that $t_{x,x}$ expresses the  single-particle potential.

Since the Hamiltonian $\hH$ commutes with the total particle number operator $\hN$, each energy eigenstate can be labeled by the eigenvalue $N$ of $\hN$. Let $\Gk$ be a normalized ground state and $\EGS_N$ be the the ground state energy with $N$ particles.  We shall abbreviate the expectation value $\bkP{\cdots}$ as $\bkG{\cdots}$.
We define the chemical potential $\mu_N$ and the charge gap $\Di_N$ at $N$ by
\eqg
\mu_N=\EGS_{N+1}-\EGS_N,\lb{mu}\\
\Di_N=\EGS_{N+1}+\EGS_{N-1}-2\EGS_N.
\lb{DE}
\eng
We can show in general that $\mu_N$ is a quantity at most of order 1 \cite{en:mu}.
It is believed that the charge gap provides a simple criterion for conductivity: the ground state is insulating if $\Di_N$ is positive and of order 1 \cite{LiebWu}.
We define the charge susceptibility at $N$ by
\eq
\chi_N=\frac{1}{\NLa}\frac{1}{\mu_N-\mu_{N-1}}=\frac{1}{\NLa\,\Di_N}.
\lb{kappa}
\en

Let $\epsilon(\rho)$ be the thermodynamic limit of the ground state energy density $\EGS_N/\NLa$ as a function of the particle density $\rho$ \cite{Ruelle}.
It generally holds that $\epsilon''(\rho)>0$ except at phase coexistence points.
If $\EGS_N/\NLa$ convergence to $\epsilon(\rho)$ in a sufficiently uniform manner, $\chi_N$ for sufficiently large but finite $\La$ should  be close to its thermodynamic limit $\chi(\rho)=1/\epsilon''(\rho)$, and hence should be positive.
We expect this to be the case for most $N$ in  a generic interacting system.
(On the other hand $\chi_N$ exhibits a pathological behavior, e.g., in a free fermion system.)
We note however that Theorem 1 below is valid no matter whether this expectation holds or not.
In section~A of \cite{SM} we directly treat $\chi(\rho)$ defined in the thermodynamic limit.

To test for possible ODLRO, we introduce the order operator
\eq
\hO=\sum_{x\in\La}\zeta_x\,\ha_x,
\lb{O}
\en
where $\zeta_x\in\bbC$ is an arbitrary constant such that $|\zeta_x|\le1$.
For any choice of $\zeta_x$, the order operator changes the particle number by 1:
\eq
[\hN,\hO]=-\hO,\quad[\hN,\hOd]=\hOd.
\lb{OO}
\en
We say that the ground state $\Gk$ exhibits ODLRO if, for an appropriate choice of $\zeta_x$, we have
\eq
\bkG{\hOd\hO}\geq\eta\NLa^2,
\lb{ODLRO}
\en
with a constant $\eta>0$.

Below we shall show an elementary but important inequality
\eq
\bkG{\,[\hO,[\hH,\hOd]]\,}\le A\NLa+BN,
\lb{ext}
\en
which we call the extensivity bound, 
where $A=\NLa^{-1}\sum_{x,y\in\La}|t_{x,y}|$ and $B=2\{\max_x U_x+\max_x\sum_y|V_{x,y}|\}$.
We shall assume that $A$ and $B$ are quantities of order 1.
Note that this is the case for any non-pathological Hamiltonian.
We finally note that the commutation relation $[\hO,\hOd]=\sum_x|\zeta_x|^2$ implies
\eq
\bigl|\bkG{\,[\hO,\hOd]\,}\bigr|\le C\NLa,
\lb{com}
\en
with $C=1$.

Our main result is the following.

\prop{Theorem}
For any choice of $\zeta_x$, we have
\eq
\Di_N \bkG{\hOd\hO}\le\{A+C|\mu_N|\}\NLa+B N.
\lb{main}
\en

To see the implication of the theorem let us assume  that the ground state $\Gk$ exhibits ODLRO as in \rlb{ODLRO}.
Then we see from \rlb{main} that 
\eq
\Di_N\le \frac{A+B\rho+C|\mu_N|}{\eta\NLa},
\lb{Dle}
\en
where $\rho=N/\NLa$ is the particle density.
Assuming that $\rho=O(1)$, we find that the charge gap is $O(|\Lambda|^{-1})$ and vanishes in the limit of large $|\Lambda|$.
(To be precise, $\Di_N$ may be negative in exceptional cases.)

Conversely if the charge gap satisfies $\Di_N\ge c\NLa^{-1+\nu}$ with some constants $c$ and $\nu>0$, then \rlb{Dle} implies $\eta=O(\NLa^{-\nu})$ and hence the ground state $\Gk$ does not exhibit ODLRO.

Assuming that $\chi_N>0$, the inequality \rlb{Dle} and \rlb{kappa} implies
\eq
\chi_N\ge \frac{\eta}{A+B\rho+C|\mu_N|}.
\lb{kge}
\en
This means that a ground state with ODLRO always has nonzero charge susceptibility, with possible exceptions in states with $\chi_N\le0$ (where $\chi_N$ does not have a physical meaning as susceptibility).
That the charge susceptibility $\chi_N$ is nonzero implies that the particle density of the ground state varies in response to the change in the chemical potential.
We thus conclude that the ground state with nonzero ODLRO is inevitably ``compressible''.

\para{Proof of Theorem}
From the variational principle we have
\eqg
\bkG{\hO\big(\hH-\EGS_{N}\big)\hOd}\ge\big(\EGS_{N+1}-\EGS_{N}\big)\bkG{\hO\hOd},\\
\bkG{\hOd\big(\hH-\EGS_{N}\big)\hO}\ge\big(\EGS_{N-1}-\EGS_{N}\big)\bkG{\hOd\hO},
\eng
where we noted that $\hO$ and $\hOd$ change the particle number by $1$.
By adding the two inequalities, we find
\eqa
\bkG{\,[\hO,[\hH,\hOd]]\,}&\ge\Di_N\,\bkG{\hOd\hO}+\mu_N\bkG{\,[\hO,\hOd]\,}.
\lb{proof2}
\ena
The expression on the left-hand side may not be obvious, but can be checked by a straightforward calculation.
We then get  \rlb{main} by recalling \rlb{ext} and \rlb{com}.

\para{Derivation of \rlb{ext}}
We see from an explicit computation that
\eq
[\ha_y,[\hH,\had_x]]=t_{y,x}+V_{x,y}\,\had_x\ha_y,
\en
for any $x,y\in\La$ such that $x\ne y$, and
\eq
[\ha_x,[\hH,\had_x]]=t_{x,x}+2U_x\hn_x+\sum_yV_{x,y}\hn_y,
\en
for any $x\in\La$.
Since $|\zeta_x|\le1$, we have
\eqa
&\bkG{\,[\hO,[\hH,\hOd]]\,}\le\sum_{x,y}\bigl|\bkG{\,[\ha_y,[\hH,\had_x]]\,}\bigr|
\nl&\le\sum_{x,y}|t_{x,y}|+2\sum_xU_x\bkG{\hn_x}+2\sum_{x,y}|V_{x,y}|\,\bkG{\hn_x},
\ena
where we noted that the Schwartz  and the arithmetic-geometric mean inequalities imply  $|\bkG{\had_x\ha_y}|\le\sqrt{\bkG{\hn_x}\bkG{\hn_y}}\le(\bkG{\hn_x}+\bkG{\hn_y})/2$.
We get \rlb{ext} by noting that $\sum_x\bkG{\hn_x}=N$.

\para{Generalization}
It is obvious from the proof that our theorem can be generalized to a much larger class of quantum particle systems and quantum spin systems.
We can treat any system with a Hamiltonian $\hH$, a conserved charge $\hN$, i.e., a self-adjoint operator such that $[\hH,\hN]=0$, and an order operator $\hO$ such that $[\hN,\hOd]=q\hOd$ with a fixed positive integer $q$.
We usually set $q=1$ or 2.
We need to confirm the extensivity bound \rlb{ext}, which is indeed the only nontrivial assumption, and the bound \rlb{com} for the commutator.
Then the theorem is proved in exactly the same manner if we redefine, with a slight abuse of notation, the chemical potential and the charge gap as $\mu_N=\EGS_{N+q}-\EGS_N$ and $\Di_N=\mu_N-\mu_{N-q}$.
The implication of the theorem for  particle systems is  essentially the same as the lattice boson systems.
We shall discuss the implication for quantum spin systems below.

In what follows we discuss extensions to specific classes of systems separately.

\para{Other lattice boson systems}
It is not difficult to extend our proof to almost any model of single-species bosons on a lattice.
Let us remark that the treatment of three-body interactions causes an extra problem in the justification of the extensivity bound \rlb{ext} since one has to control, e.g., the summation $\sum_x\bkG{\hn_x^2}$, which can be large.
In such a case one needs to impose extra condition about the (near) translation invariance to justify \rlb{ext}.

The extension of the theorem to models of multi species bosons and/or spinfull bosons is automatic.

\para{Lattice fermion systems}
The treatment of lattice fermion systems is 
easier than bosonic systems since all operators are bounded.
It is most natural (but not mandatory) to set $\hN$ to be the total number operator.
We can treat any Hamiltonian $\hH=\sum_{x\in\La}\hh_x$  and order operator $\hO=\sum_{x\in\La}\ho_x$.
We assume, for any $x\in\La$, that $\norms{\hh_x}\le h_0$ and $\norms{\ho_x}\le o_0$ with constants $h_0$ and $o_0$, that the commutator $[\hh_x,\ho_y]$ is nonzero for at most $n$ distinct $y$'s, and that  $[\ho_x,\ho_y]$ is nonzero for at most $n'$ distinct $y$'s.
Then the extentiviy condition \rlb{ext} is readily justified for any state by only examining the operator norms, where we can take $A=8nn'h_0(o_0)^2$ and $B=0$.
The bound \rlb{com} also holds with $C=2n'(o_0)^2$.

As a typical example, we can treat the Hubbard model with short range hopping, and examine the condensation of local Cooper pairs by setting $\ho_x=\hat{c}_{x,\up}\hat{c}_{x,\dn}$, in which case we have $q=2$.

\para{Continuum particle systems}
Since our proof only makes use of the variational principle, it readily extends to continuum particle systems without losing mathematical rigor.
The only nontrivial issue is the justification of the extensivity bound \rlb{ext}.

Let us discuss a typical bosonic system.
Consider a system of bosons in a cubic region $\La\subset\bbR^3$ with volume $\NLa$.
We denote by $\hpsi(\bsr)$ and $\hpsid(\bsr)$ the standard annihilation and creation operators at $\bsr$, which satisfies $[\hpsi(\bsr),\hpsid(\bss)]=\delta(\bsr-\bss)$.
(Note that we are using physicists' notation only for a book-keeping purpose.  The following estimate can be done in a mathematically rigorous manner.)
We consider a general Hamiltonian
\eqa
\hH&=\int d^3\bsr\,\hpsid(\bsr)\Bigl\{-\frac{\Lap}{2m}+V(\bsr)\Bigr\}\hpsi(\bsr)
\nl&+\frac{1}{2}\int d^3\bsr\,d^3\bss\,\hpsid(\bsr)\hpsid(\bss)\,V_{\rm int}(\bsr,\bss)\,\hpsi(\bss)\hpsi(\bsr),
\ena
with single-body potential $V(\bsr)$ and interaction potential $V_{\rm int}(\bsr,\bss)=V_{\rm int}(\bss,\bsr)$.

Take a complex valued function $\zeta(\bsr)$ such that $|\zeta(\bsr)|\le1$ and $|\Lap \zeta(\bsr)|\le\bar{p}^2$ for any $\bsr$, where $\bar{p}>0$ is a constant.
We define the corresponding order operator by $\hO=\int d^3\bsr\,\zeta^*(\bsr)\hpsi(\bsr)$, which satisfies \rlb{com} with $C=1$.
A straightforward calculation shows 
\eqa
[\hO,[\hH,\hOd]]&=\int d^3\bsr\,\zeta^*(\bsr)\Bigl\{-\frac{\Lap}{2m}+V(\bsr)\Bigr\}\zeta(\bsr)
\nl&+\int d^3\bsr\,d^3\bss\,|\zeta(\bsr)|^2\,V_{\rm int}(\bsr,\bss)\,\hpsid(\bss)\hpsi(\bss)
\nl&+\int d^3\bsr\,d^3\bss\,\zeta^*(\bsr)\zeta(\bss)\,V_{\rm int}(\bsr,\bss)\,\hpsid(\bss)\hpsi(\bsr).
\ena
As in the lattice model, we bound the expectation value as
\eqa
\bkG{\,[\hO,[\hH,\hOd]]\,}
&\le\int d^3\bsr\Bigl\{\frac{\bar{p}^2}{2m}+|V(\bsr)|\Bigr\}
\nl&+2\int d^3\bsr\,d^3\bss\,|V_{\rm int}(\bsr,\bss)|\,\rho(\bss),
\ena
where $\rho(\bss)=\bkG{\hpsid(\bss)\hpsi(\bss)}$ is the particle density in the ground state $\Gk$.
This implies the desired \rlb{ext} with
\eqa
A&=\frac{\bar{p}^2}{2m}+\frac{1}{\NLa}\int d^3\bsr\,|V(\bsr)|,\\
B&=2\sup_{\bsr}\int d^3\bss\,|V_{\rm int}(\bsr,\bss)|.
\ena

\para{Quantum spin systems}
The treatment of quantum spin systems is also easy since all operators are bounded.
Let us consider any Hamiltonian $\hH$ (satisfying the same conditions as in the fermion systems) that commutes with $\hN=\sum_x\hS^{(3)}_x$.
We here denoted  the spin operator as $(\hS^{(1)}_x,\hS^{(2)}_x,\hS^{(3)}_x)$ and set $\hS^-_x=\hS^{(1)}_x-i\hS^{(2)}_x$.
We set  the order operator as $\hO=\sum_x\zeta_x\hS^-_x$ with $\zeta_x\in\bbR$ such that $|\zeta_x|\le1$, where we have $q=1$.
The conditions \rlb{ext} and \rlb{com} are clearly satisfied.
Since
\eq
\hOd\hO+\hO\hOd=2\sum_{x,y\in\La}\zeta_x\zeta_y(\hS^{(1)}_x\hS^{(1)}_y+\hS^{(2)}_x\hS^{(2)}_y),
\en
we see that $\bkG{\hOd\hO}$ is a measure of long-range order in the 1 and 2 directions in the ground state $\Gk$ with fixed $N$, i.e., the total spin in the 3-direction.

In a quantum spin system, it is standard (and physically realistic) to consider the whole Hilbert space without restricting the eigenvalue of $\hN$, and investigate the ground state of the Hamiltonian $\hH_h=\hH-h\hN$, where $h$ is the uniform magnetic field.
Suppose that $\Gk$ is a ground state of $\hH_h$ for some $h$.
We define $h_+=\min_{n>0}(\EGS_{N+n}-\EGS_N)/n$ and $h_-=\max_{n>0}(\EGS_N-\EGS_{N-n})/n$.
One then finds that $\Gk$ is the unique ground state of $\hH_h$ for $h\in(h_-,h_+)$.
We thus see that $\tilde{\Di}_N=h_+-h_-$ is the width of the magnetization plateau in the magnetization process where $h$ is varied.
Note that $0\le\tilde{\Di}_N\le\Di_N$ by definition.

Suppose that $\tilde{\Di}_N>0$ is of order 1, in other words, that the ground state $\Gk$ belongs to a magnetization plateau.
Then \rlb{Dle} implies that $\eta=O(|\La|^{-1})$, i.e.,  $\Gk$ has no long-range order in the directions perpendicular to the uniform magnetic field.
We can similarly rule out long-range order corresponding to general $q>1$.  Notable examples are the nematic and triatic order, which correspond to $q=2$ and $q=3$ \cite{ChandraColeman,Chubukov,MomoiSindzingreShannon,KeckeMomoiFurusaki}.

See section~A of \cite{SM} for the implication of our theory for ground states with long-range order.

\para{Existing results for symmetry-breaking state}
Here, focusing on bosonic systems, we review a bound on susceptibility derived in \cite{Wagner1966,Stringari} that applies to a ground state that breaks the U(1) symmetry, and clarify the relation to our bound \rlb{kge}.

So far we examined properties of the ground state $\Gk$ that exhibits ODLRO as in \rlb{ODLRO}. 
Note that  $\Gk$, which is also an eigenstate of the charge $\hN$, never breaks the U(1) symmetry, as manifested in the fact that $\bkGS{\hO}_N=0$. To have a U(1) symmetry breaking ground state, we need to consider the whole Hilbert space without restricting the eigenvalue of $\hN$, and take a superposition of states with different particle numbers. Such a superposition is a purely theoretical object and is never realized (or, to be more precise, never observed) in a system of massive particles.

To this end, we consider the Hamiltonian
\eq
\hH_{\mu,\sbf}=\hH-\mu\hN-\sbf(\hO+\hOd),
\lb{Hmemain}
\en
where $\hH$ is the U(1) symmetric Hamiltonian such as  \rlb{H}, $\mu$ is the chemical potential as an external control parameter, and $\sbf$ is the symmetry breaking field.
Note that the symmetry breaking term is strictly artificial in a particle system since it violates the particle number conservation law.
(In a quantum spin system, this term may be realized by external (staggered) magnetic filed in the 1-direction.)
We denote by $\tGk$ the ground state of the Hamiltonian \rlb{Hmemain}.

Since the ground state $\tGk$ is not an eigenstate of $\hat{N}$, we focus on the expectation value $\tNGS_{\mu,\sbf}=\tGb\hN\tGk$, and define the charge susceptibility by 
\eq
\tilde{\chi}(\mu,\sbf)=\frac{1}{|\La|}\frac{\partial}{\partial\mu}\tNGS_{\mu,\sbf}.
\lb{CSmain}
\en
As shown in \cite{Wagner1966,Stringari}, the Bogoliubov inequality for $\hO$ and $\hN$ implies
\eq
\tilde{\chi}(\mu,\sbf)\ge\frac{\bigl|\tGb\hO\tGk\bigr|^2}{|\La|\tGb\,[\hO,[\hH_{\mu,\sbf},\hOd]]\,\tGk}.
\lb{Wagnermain}
\en
See section B of \cite{SM} for a proof.

Let us evaluate the right-hand side of \rlb{Wagnermain}. 
For a fixed $\mu$, suppose that $\Gk$ is a ground state (within the whole Hilbert space) of $\hH_{\mu,0}$, and that $\Gk$ satisfies the conditions \rlb{ODLRO}, \rlb{ext}, and \rlb{com}. 
Then, it has been proved that
\eq
|\tGb\hO\tGk|\gtrsim\sqrt{\eta}\,|\La|,
\lb{OLBmain}
\en
provided that $\lambda |\La|^2$ is large enough \cite{HorschLinden,KaplanHorschLinden,Tasaki2017,TasakiBook}. See section~B of \cite{SM} for details.  For the denominator in \rlb{Wagnermain}, one easily verifies an extensivity bound
\eq
\tGb\,[\hO,[\hH_{\mu,\sbf},\hOd]]\,\tGk\le (A+C\mu)|\La|+B \tNGS_{\mu,\sbf}.
\lb{ex2main}
\en
By substituting \rlb{OLBmain} and \rlb{ex2main} into \rlb{Wagnermain}, we find
\eq
\tilde{\chi}(\mu,\sbf)\gtrsim\frac{\eta}{A+B\tilde{\rho}+C\mu},
\lb{Wagner2main}
\en
where $\tilde{\rho}=\tNGS_{\mu,\sbf}/|\Lambda|$ is the particle density of the symmetry breaking state $\tGk$, which should coincide with  $\rho=N/|\Lambda|$ in the limit of large $|\Lambda|$.
The bound \rlb{Wagner2main} is, at least apparently, the same as our bound \rlb{kge}. 

Although one might be tempted to conclude that our inequality \rlb{kge} is a re-derivation of the above result, this is not the case.
Note that we have $\tilde\chi(\mu,0)=0$ in any finite system because of the U(1) symmetry.
To get a meaningful consequence \rlb{Wagner2main} from \rlb{Wagnermain}, we must consider a system with sufficiently large symmmetry breaking external field $\sbf>0$, which is indeed strictly prohibited in nature. 
Our bound \rlb{kge}, on the contrary, is formulated and proved in a physically meaningful setting with a particle number conserving Hamiltonian and the Hilbert space with fixed $N$.
Our definition \rlb{kappa} of the charge susceptibility represents a quantity that can be measured in principle in experiments, and are indeed calculated in numerical simulations.

With sufficient knowledge about symmetry breaking, one can extract from the inequality \rlb{Wagner2main} essentially the same physical implication as our  inequality \rlb{kge}.
This is because the U(1) symmetry breaking ground state $\tGk$ with suitably small $\sbf$ is believe to be  (close to) a superposition of the ground states $\Gk$, and, moreover, the values of $N$ are essentially concentrated in a small interval \cite{TasakiBook}.
This suggests that, as far as one is concerned with thermodynamic properties described by the energy density $\epsilon(\rho)$, the symmetry breaking ground state $\tGk$ and the symmetry preserving ground state $\Gk$ exhibit the same properties.

\para{Discussion}%
For a general class of quantum many-body systems, we proved a simple inequality which states that a ground state with ODLRO inevitably has a vanishing charge gap, or, almost equivalently, nonzero charge susceptibility $\chi$, and hence is ``compressible''.
As we noted in the introduction, it is interesting to apply this conclusion to a commensurate supersolid in lattice boson systems, where ODLRO is observed at a commensurate filling that is consistent with a certain crystalline ordering of particles \cite{OhgoeSuzukiKawashima2012,SuzukiKoga2014}.
The ground state is incompressible, i.e., $\chi=0$, when it is in the Mott insulating phase, but should become compressible, i.e., $\chi>0$, once it enters the supersolid phase where ODLRO coexists with a spontaneous breakdown of the translation symmetry.
This rather surprising consequence of our theorem is indeed consistent with numerical observations in \cite{OhgoeSuzukiKawashima2012,SuzukiKoga2014}.
Essentially the same behavior is expected in a recently proposed exactly solvable model of interacting bosons, where the ground state resembles that of a Mott insulator \cite{KKMTT2021}.

Let us finally make a brief comment on the relation between our finding and the Goldstone theorem \cite{Wagner1966,Stringari,Momoi94,Momoi95,Koma2021}.
Although both our theory and the Goldstone theorem are about symmetry breaking (or ODLRO) and energy gaps, we should say that the two are essentially different.
First, when the system has translation invariance, our theory compares the energies of two states with zero momentum, while the Goldstone theorem compares the energies of the zero momentum ground state and an excited state with nonzero momentum.
(In fact, our theory applies to systems which do not have any translation invariance.)
Secondly, and more importantly, our theory deals with the energy gap between two U(1) invariant ground states (with different fixed particle numbers), while the Goldstone theorem deals with the energy gap between a ground state with broken U(1) symmetry and an excited state above it.

\medskip
{\small
It is a pleasure to thank David Huse for a valuable exchange that triggered us to work on the present problem, and 
Hosho Katsura, Naoki Kawashima, Tohru Koma, Kenn Kubo, Elliott Lieb, Tsutomu Momoi, Masaki Oshikawa, Akinori Tanaka, and Masafumi Udagawa for valuable discussions.
H.W. was supported by JSPS Grant-in-Aid for Scientific Research No.~JP20H01825 and by JST PRESTO Grant No.~JPMJPR18LA. 
}



\clearpage

\setcounter{page}{1}

\onecolumngrid

\begin{center}
{\bf \large Supplemental Material for ``Off-Diagonal Long-Range Order Implies Vanishing Charge Gap"}
\end{center}

\bigskip\noindent
{\bf \large A. The inequality for the charge susceptibility in the thermodynamic limit}
\setcounter{equation}{0}
\def\theequation{A.\arabic{equation}}
\setcounter{figure}{0}
\def\thefigure{A.\arabic{figure}}

\medskip
We shall state and prove a result corresponding to \rlb{kge} for the charge susceptibility $\chi(\rho)$ defined in the thermodynamic limit.
For notational simplicity, we shall restrict ourselves to the case with $q=1$, with bosonic systems in mind.
Generalization to $q>1$ is straightforward.
The results are valid both for lattice systems and continuum systems.

Let $\La_j$ with $j=1,2,\ldots$ be a sequence of finite lattices (or finite regions in $\bbR^3$) such that $|\La_j|\up\infty$ as $j\up\infty$.
We consider a system on $\La_j$ for each $j$, and label various quantities explicitly by the index $j$ as $\EGS_{j,N}$, $\mu_{j,N}$, or $\bkGS{\cdots}_{\!j,N}$.
For a given density $\rho$, we denote by $N_j$ the integer closest to $\rho|\La_j|$.
We assume that for any $\rho$ (in the range allowed by the definition of the model) the limit
\eq
\eGS(\rho)=\lim_{j\up\infty}\frac{\EGS_{j,N_j}}{|\La_j|}
\en
exists, and is convex in $\rho$.
This fact is proved in general if $\La_j$ is a sequence of hypercubic lattices (or regions) and the Hamiltonians $\hH_j$ have suitable translation invariance \cite{Ruelle}.
By convexity we can define the chemical potential $\mu(\rho)$ as the right (or left) derivative of $\eGS(\rho)$.

We further assume that there are constants, $\rho_0$, $\rho_1$, $\mu_0$, and $\eta_0$, such that
\eq
|\mu_{j,N}|\le\mu_0,\quad\bkGS{\hOd\hO}_{\!j,N}\ge\eta_0\,|\La_j|^2
\lb{ass2}
\en
hold for any $j$ and $N$ such that $\rho_0\le N/|\La_j|\le\rho_1$.
Note that the second assumption in \rlb{ass2}, which requires all the ground states in the density range to exhibit ODLRO, is rather strong,
while the first assumption is easily verified in many systems.

\para{Theorem A.1}
Under the above assumption, we have, for any $\rho,\rho'\in(\rho_0,\rho_1)$ with $\rho\ne\rho'$, that
\eq
0\le\frac{\mu(\rho)-\mu(\rho')}{\rho-\rho'}\le\frac{A+B\eta_0+C\rho_1}{\eta_0}.
\lb{main2}
\en

\medskip

Assuming that $\eGS(\rho)$ is twice differentiable, we see from \rlb{main2} that
\eq
\chi(\rho)=\frac{1}{\eGS''(\rho)}\ge\frac{\eta_0}{A+B\eta_0+C\rho_1}
\lb{chiTDL}
\en
We have thus obtained a bound for $\chi(\rho)$ as a  thermodynamic quantity.

Note that, in a quantum spin system, $\chi(\rho)$ is nothing but the standard magnetic susceptibility.
We have proved that the existence of transverse long-range order implies that the susceptibility is nonzero.

\para{Proof of Theorem A.1}
The first inequality in \rlb{main2} readily follows from the convexity.
From the definition \rlb{DE}, we have
\eq
\sum_{Q=0}^{M-1}\sum_{N=K+Q+1}^{L+Q}\Di_{j,N}=(\EGS_{j,L+M}-\EGS_{j,L})-(\EGS_{j,K+M}-\EGS_{j,K}).
\en
Since \rlb{Dle} implies $\Di_{j,N}\le(A+B\eta_0+C\rho_1)/(\eta_0|\La_j|)$, we see that
\eq
\frac{|\La_j|}{L-K}\frac{|\La_j|}{M}\frac{1}{|\La_j|}\bigl\{(\EGS_{j,L+M}-\EGS_{j,L})-(\EGS_{j,K+M}-\EGS_{j,K})\bigr\}
\le\frac{A+B\eta_0+C\rho_1}{\eta_0}.
\en
Taking the $j\up\infty$ limit, this implies
\eq
\frac{1}{(\rho-\rho')\,\delta}\bigl[\{\eGS(\rho+\delta)-\eGS(\rho)\bigr\}-\bigl\{\eGS(\rho'+\delta)-\eGS(\rho')\}\bigr]
\le\frac{A+B\eta_0+C\rho_1}{\eta_0},
\en
where $\rho+\delta,\rho'+\delta\in(\rho_0,\rho_1)$.
We get \rlb{main2} by letting $\delta\dn0$.~\qedm

\bigskip\noindent
{\bf \large B. Ground states that break the U(1) symmetry}
\setcounter{equation}{0}
\def\theequation{B.\arabic{equation}}

\medskip

Here we shall review some results about ground states that break the U(1) symmetry.
As we have already stressed in the main text, such a state inevitably does not have a fixed particle number.

We mainly focus on bosonic systems, and always work within the whole Hilbert space, namely, the Fock space, without restricting the particle number $N$.
All the results are valid both for lattice systems and continuum systems.
We fix a lattice (or a finite region in $\bbR^3$) $\La$ unless specified.

We consider the Hamiltonian
\eq
\hH_{\mu,\sbf}=\hH-\mu\hN-\sbf(\hO+\hOd),
\lb{Hme}
\en
which is the same as  \rlb{Hmemain} in the main text.

\bigskip
{\bf The relation between off-diagonal long-range order and symmetry breaking:}
Let us first discuss a theorem proved by Horsch, and von~der~Linden \cite{HorschLinden} and Kaplan, Horsch, and von~der~Linden \cite{KaplanHorschLinden} that clarifies the relation between off-diagonal long-range order (ODLRO) and spontaneous symmetry breaking.

As in the main text, we denote by $\Gk$ a ground state of $\hH$ with a fixed particle number $N$.
Since $\hH_{\mu,0}$ commute with $\hN$, one finds that $\Gk$ with a suitable $N$ is  a ground state of $\hH_{\mu,0}$.
Then the following theorem is proved in \cite{HorschLinden} and \cite{KaplanHorschLinden}.

\para{Theorem B.1}
For a fixed $\mu$, let $\Gk$ be a ground state (within the large Hilbert space) of $\hH_{\mu,0}$.
Assume that $\Gk$ satisfies the conditions \rlb{ODLRO}, \rlb{ext}, and \rlb{com}.
Then for any $\sbf>0$, it holds that
\eq
\frac{1}{|\La|}\tGb(\hO+\hOd)\tGk\ge\sqrt{2\eta-C/|\La|}-\frac{A+B\rho+C\mu}{2(2\eta-C/|\La|)}\frac{1}{\sbf\,|\La|^2},
\lb{OOLB}
\en
where $\tGk$ is an arbitrary ground state of $\hH_{\mu,\sbf}$.

\medskip
The most essential assumption here is  \rlb{ODLRO} about the existence of ODLRO.
The inequality \rlb{OOLB} shows that
\eq
\frac{1}{|\La|}\tGb(\hO+\hOd)\tGk\gtrsim\sqrt{2\eta},
\lb{OOLB2}
\en
provided that $|\La|$ is large enough and also $\sbf>0$ is not too small to guarantee that $\sbf\,|\La|^2$ is large enough.
In fact one can use a more sophisticated variational state proposed in \cite{KomaTasaki1994} to prove a better (and optimal) bound
\eq
\frac{1}{|\La|}\tGb(\hO+\hOd)\tGk\gtrsim2\sqrt{\eta}.
\en
See \cite{Tasaki2017,TasakiBook}.
Noting that the real part of $\tGb\hO\tGk$ is $\tGb(\hO+\hOd)\tGk/2$, this leads to the lower bound \rlb{OLBmain} in the main text.

Note that the inequality \rlb{OOLB} is meaningless if we let $\sbf\dn0$ in a finite system since the right-hand side becomes negative.
This is inevitable since there is no symmetry breaking in a finite system with $\sbf=0$.
To see spontaneous symmetry breaking one needs to first take the infinite volume limit, and then let $\sbf\dn0$.
The following is a straightforward corollary of the theorem.

\para{Corollary B.2}
Suppose that the assumption of Theorem~B.1 is satisfied, again for a fixed $\mu$, for a sequence of lattices (or regions in $\bbR^3$) $\La_j$ (with $j=1,2,\ldots$) such that $|\La_j|\up\infty$ as $j\up\infty$.
Then the ground states exhibit spontaneous breaking of the U(1) symmetry in the sense that
\eq
\liminf_{\sbf\dn0}\,\liminf_{j\up\infty}\frac{1}{|\La|}\tGb(\hO+\hOd)\tGk\ge\sqrt{2\eta}.
\lb{B:SSB}
\en

\medskip
Again the right-hand side can be replaced by $2\sqrt{\eta}$ if we use the improved bound in \cite{Tasaki2017,TasakiBook}.

\para{Proof of Theorem~B.1}
We write the ground state energy of $\hH_{\mu,0}$ as $\tEGS_{\mu,0}=\EGS_N-\mu N$.
Following \cite{HorschLinden}, we define a normalized trial state 
\eq
\ket{\Xi}=\frac{1}{\sqrt{2}}\biggl\{\Gk+\frac{(\hO+\hOd)\Gk}{\sqrt{\bkG{\hOd\hO}+\bkG{\hO\hOd}}}\biggr\}.
\en
An explicit calculation shows
\eqa
\bra{\Xi}\hH_{\mu,0}\ket{\Xi}&=\tEGS_{\mu,0}+\frac{1}{2}\frac{\bbkG{(\hO+\hOd)(\hH_{\mu,0}-\tEGS_{\mu,0})(\hO+\hOd)}}{\bkG{\hOd\hO}+\bkG{\hO\hOd}}
\nl&=\tEGS_{\mu,0}+\frac{1}{2}\frac{\bbkG{\,[\hO,[\hH_{\mu,0},\hOd]]\,}}{\bkG{\hOd\hO}+\bkG{\hO\hOd}}
\nl&=\tEGS_{\mu,0}+\frac{1}{2}\frac{\bbkG{\,[\hO,[\hH,\hOd]]\,}-\mu\bkG{\,[\hO,\hOd]\,}}{2\bkG{\hOd\hO}+\bkG{\,[\hO,\hOd]\,}}
\nl&\le
\tEGS_{\mu,0}+\frac{1}{2}\frac{A|\La|+BN+C\mu|\La|}{2\eta|\La|^2-C|\La|}
\nl&=
\tEGS_{\mu,0}+\frac{A+B\rho+C\mu}{2(2\eta-C/|\La|)}\,\frac{1}{|\La|},
\lb{XEX}
\ena
where we used  \rlb{ODLRO}, \rlb{ext}, and \rlb{com}.
We also see that
\eqa
\bra{\Xi}(\hO+\hOd)\ket{\Xi}&=\frac{\bbkG{(\hO+\hOd)^2}}{\sqrt{\bkG{\hOd\hO}+\bkG{\hO\hOd}}}\nl&=\sqrt{\bkG{\hOd\hO}+\bkG{\hO\hOd}}\nl&\ge\sqrt{2\eta|\La|^2-C|\La|}.
\lb{XOOX}
\ena

Note that the variational principle implies
\eq
\tGb\hH_{\mu,\sbf}\tGk\le\bra{\Xi}\hH_{\mu,\sbf}\ket{\Xi}.
\en
By recalling the definition \rlb{Hme} of the Hamiltonian, we follow \cite{KaplanHorschLinden} and rearrange the inequality as
\eqa
\frac{1}{|\La|}\tGb(\hO+\hOd)\tGk&\ge\frac{1}{|\La|}\bra{\Xi}(\hO+\hOd)\ket{\Xi}
+\frac{1}{\sbf\,|\La|}\Bigl\{\tGb\hH_{\mu,0}\tGk-\bra{\Xi}\hH_{\mu,0}\ket{\Xi}\Bigr\}
\nl&\ge\sqrt{2\eta-C/|\La|}-\frac{A+B\rho+C\mu}{2(2\eta-C/|\La|)}\frac{1}{\sbf\,|\La|^2},
\ena
where we noted that $\tGb\hH_{\mu,0}\tGk\ge\tEGS_{\mu,0}$ and used the bounds \rlb{XEX} and \rlb{XOOX}.~\qedm

\bigskip
{\bf Bogoliubov inequality for the charge susceptibility}:
We shall now discuss the inequality \rlb{Wagnermain}, which has very similar implication as our inequality \rlb{kge} for the charge susceptibility.

Suppose that, for fixed $\mu$ and $\sbf>0$, the Hamiltonian $\hH_{\mu,\sbf}$ has a unique ground state $\tGk$.
For notational simplicity, we fix $\mu$ and $\sbf$, and write the energy eigenstates of $\hH_{\mu,\sbf}$ as $\ket{\Psi_j}$, where $j=0,1,2,\ldots$, with $\hH_{\mu,\sbf}\ket{\Psi_j}=\tilde{E}_j\ket{\Psi_j}$.
We set $\ket{\Psi_0}=\tGk$.
As in \rlb{CSmain} in the main text, we define the charge susceptibility by
\eq
\tilde{\chi}(\mu,\sbf)=\frac{1}{|\La|}\frac{\partial}{\partial\mu}\tGb\hN\tGk=\frac{2}{|\La|}\sum_{j\ne0}\bra{\Psi_0}\hN\ket{\Psi_j}\frac{1}{E'_j}\bra{\Psi_j}\hN\ket{\Psi_0},
\en 
where we set $E'_j=\tilde{E}_j-\tilde{E}_0>0$.
The final expression follows from the standard Rayleigh-Schr\"odinger perturbation theory.

Then the following inequality was proved in  \cite{Wagner1966,Stringari}.

\para{Theorem B.3}
Assume that $\hH_{\mu,\sbf}$ has a unique ground state $\tGk$.
Then the above defined susceptibility satisfies
\eq
\tilde{\chi}(\mu,\sbf)\ge\frac{\bigl|\tGb\hO\tGk\bigr|^2}{|\La|\,\tGb\,[\hO,[\hH_{\mu,\sbf},\hOd]]\,\tGk}.
\lb{Wagner}
\en

\para{Proof of Theorem B.3}
We first observe that
\eqa
\bra{\Psi_0}\,[\hO,\hN]\,\ket{\Psi_0}&=
\sum_{j\ne0}\Bigl\{
\bra{\Psi_0}\hO\ket{\Psi_j}\bra{\Psi_j}\hN\ket{\Psi_0}-\bra{\Psi_0}\hN\ket{\Psi_j}\bra{\Psi_j}\hO\ket{\Psi_0}
\Bigr\}
\nl&=\sum_{j\ne0}\bra{\Psi_0}\hO\ket{\Psi_j}\sqrt{\smash[b]{E'_j}}\frac{1}{\sqrt{\smash[b]{E'_j}}}\bra{\Psi_j}\hN\ket{\Psi_0}
-\sum_{j\ne0}\bra{\Psi_0}\hN\ket{\Psi_j}\frac{1}{\sqrt{\smash[b]{E'_j}}}\sqrt{\smash[b]{E'_j}}\bra{\Psi_j}\hO\ket{\Psi_0},
\ena
which implies
\eq
\bigl|\bra{\Psi_0}\,[\hO,\hN]\,\ket{\Psi_0}\bigr|^2
\le2\Biggl\{
\biggl|\sum_{j\ne0}\bra{\Psi_0}\hO\ket{\Psi_j}\sqrt{\smash[b]{E'_j}}\frac{1}{\sqrt{\smash[b]{E'_j}}}\bra{\Psi_j}\hN\ket{\Psi_0}\biggr|^2
+
\biggl|\sum_{j\ne0}\bra{\Psi_0}\hN\ket{\Psi_j}\frac{1}{\sqrt{\smash[b]{E'_j}}}\sqrt{\smash[b]{E'_j}}\bra{\Psi_j}\hO\ket{\Psi_0}\biggr|^2
\Biggr\}.
\lb{0}
\en
Then we note from the Schwartz inequality that
\eqa
\biggl|\sum_{j\ne0}\bra{\Psi_0}\hO\ket{\Psi_j}\sqrt{\smash[b]{E'_j}}\frac{1}{\sqrt{\smash[b]{E'_j}}}\bra{\Psi_j}\hat{N}\ket{\Psi_0}\biggr|^2
&\le\Bigl(\sum_{j=0}^\infty\bra{\Psi_0}\hO\ket{\Psi_j}E'_j\bra{\Psi_j}\hOd\ket{\Psi_0}\Bigr)
\Bigl(\sum_{j\ne0}\bra{\Psi_0}\hN\ket{\Psi_j}\frac{1}{E'_j}\bra{\Psi_j}\hN\ket{\Psi_0}\Bigr)
\nl&=\bra{\Psi_0}\hO(\hH_{\mu,\sbf}-\tEGS_{\mu,\sbf})\hOd\ket{\Psi_0}\,\frac{\tilde{\chi}(\mu,\sbf)\,|\La|}{2},
\lb{1}
\ena
and also that
\eq
\biggl|\sum_{j\ne0}\bra{\Psi_0}\hN\ket{\Psi_j}\frac{1}{\sqrt{\smash[b]{E'_j}}}\sqrt{\smash[b]{E'_j}}\bra{\Psi_j}\hO\ket{\Psi_0}\biggr|^2\le \bra{\Psi_0}\hOd(\hH_{\mu,\sbf}-\tEGS_{\mu,\sbf})\hO\ket{\Psi_0}\,\frac{\tilde{\chi}(\mu,\sbf)\,|\La|}{2}.
\lb{2}
\en
Substituting \rlb{1} and \rlb{2} into \rlb{0}, we get
\eq
\bigl|\bra{\Psi_0}\,[\hO,\hN]\,\ket{\Psi_0}\bigr|^2\le\bra{\Psi_0}\,[\hO,[\hH_{\mu,\sbf},\hOd]]\,\ket{\Psi_0}\,\tilde{\chi}(\mu,\sbf)\,|\La|,
\en
which is sometimes called the Bogoliubov inequality.
The desired inequality \rlb{Wagner} follows since $[\hO,\hN]=\hO$.~\qedm

\end{document}